\documentclass[%
reprint,
 aps,
 pre,%
 amsmath,amssymb,
]{revtex4-1}

\usepackage{graphicx}
\usepackage{epstopdf}
\usepackage{dcolumn}
\usepackage{bm}
\usepackage{subfig}
\usepackage{color}

\begin{document}


\title{Spinodal decomposition of a binary magnetic fluid confined to a surface}


\author{K. Lichtner}
\affiliation{%
Institute of Theoretical Physics, Secr. EW~7-1, Technical University Berlin, \\Hardenbergstr. 36, D-10623 Berlin, Germany
}%
\email{lichtner@mail.tu-berlin.de}

\author{S. H. L. Klapp}
\affiliation{%
Institute of Theoretical Physics, Secr. EW~7-1, Technical University Berlin, \\Hardenbergstr. 36, D-10623 Berlin, Germany
}%

\date{\today}

\begin{abstract}
%
In our previous work [J. Chem. Phys. \textbf{136}, 024502 (2012)], we reported a demixing phase transition of a quasi-two-dimensional (2D), binary Heisenberg fluid mixture driven by the  ferromagnetic interactions of the magnetic species.
Here, we present a theoretical study for the \textit{time-dependent} coarsening occuring within the two-phase region in the density-concentration plane, also known as spinodal decomposition. 
Our investigations are based on Dynamical Density Functional Theory (DDFT). 
The particles in the mixture are modelled as Gaussian soft spheres on a two-dimensional surface, where one component carries a classical spin of Heisenberg type.
To investigate the two-phase region, we first present a linear stability analysis with respect to small, harmonic density perturbations. Second, to capture non-linear 
effects, we calculate time-dependent structure factors by combining DDFT with Percus' test particle method.
For the 
growth of the average domain size $l$ during spinodal decomposition with time $t$, we observe a power-law behavior $l\propto t^{\delta_\alpha}$ with $\delta_m\simeq 0.333$ for the magnetic species and $\delta_n\simeq 0.323$ for the non-magnetic species. 
\end{abstract}

\pacs{Valid PACS appear here}
\maketitle

%



\section{Introduction}
Spinodal decomposition is a non-equilibrium phenomenon occurring in a variety of hard and soft condensed matter systems undergoing first-order phase transitions. 
It refers to the spontaneous formation of domains, i.e. spatial symmetry breaking, following a sudden quench into a two-phase region. Examples for hard materials 
displaying spinodal decomposition are binary alloys, e.g., Fe-Cr\cite{Miller19953385} where the atoms phase seperate into a two-phase microstructure, or 
metallic glasses\cite{Kim2013} where the structural properties are intriguing for practical applications\cite{PhysRevFocus.15.20}. 
In soft matter, non-trivial phase behaviour may be found, e.g., in colloidal gels\cite{DhontGel}, glasses\cite{collrev2013} and mixtures\cite{aarts2012,doi:10.1021/la048737f,PhysRevE.64.041501}, or in a more general sense in fluids with competing interactions\cite{PhysRevLett.105.045001,bomont:164901,linse:3917}.
Even more complex behavior is observed for particles with internal degrees of freedom such as magnetic particles\cite{lichtner:024502,PhysRevLett.104.255703} and shape-anisotropic particles such as colloidal rods\cite{PhysRevLett.106.208302}. \\
Theoretically, spinodal decomposition has been intensely investigated over the past decades starting from the generalized diffusion equation suggested by Cahn and Hilliard (CH)\cite{cahn:258,CahnHilliard,Cahn1961795}. 
However, since the CH theory is linear in the density perturbations, it focusses only on the early stages of phase separation kinetics\cite{dhont:5112}. Spinodal decomposition has later been 
investigated based on computer simulations\cite{binder1998spinodal,winkler:054901}, Smoluchowski equation approach\cite{dhont:5112,dhont1996introduction}, and Dynamical Density Functional Theory (DDFT)\cite{ArcherSpinDec}.\\
In the present paper we investigate a quasi-2D colloidal fluid mixture, where for one species the soft repulsive interactions between the particles are supplemented by an additional ferromagnetic interaction. 
A possible experimental realization of soft magnetic particles are complex dusty plasmas\cite{dust11,dust98}. By charging paramagnetic dust grains the magnetic interaction between the particles can be  
superimposed by a (repulsive) screened electrostatic interaction. In contrast to experiments with (2D) magnetic hard spheres\cite{MaretPRL99}, the ``softness'' of the repulsion may thus be changed by the electrostatic coupling between the grains.\\
Following an earlier study by us\cite{lichtner:024502}, we have considered the first-order demixing transition which is coupled to a ferromagnetic transition. 
In Ref.~\onlinecite{lichtner:024502} we focused on the formation of two-dimensional clusters of a (globally stable) phase surrounded by the (metastable) bulk phase.
This non-equilibrium pattern formation within the {\it metastable} regions of the phase diagram is also referred to as nucleation.\cite{ArcherEvansNucleation,Lutsko_JCP_2009_1,lichtner:024502}\\
In the present study we go one step further and investigate states deep inside the two-phase region far away of the two-phase coexistence (the binodal) 
and from the spinodal (i.e. the line of points where the barrier for nucleation vanishes). 
Inside the spinodal region of the phase diagram density perturbations with certain wavenumbers $k$ grow over time \textit{regardless} of the amplitude. 
This eventually results in a demixing process, that is,  
spinodal decomposition.\cite{ArcherSpinDec, PASTM11} \\
Our investigations are based on DDFT\cite{marconi:a413,marconi:8032,ArcherSpinDec,espanol:244101}, a generalized diffusion equation where the microscopic interactions enter via the (Helmholtz) free energy. 
Recently, DDFT has been applied to a variety of phase-separating systems including colloids with critical Casimir forces\cite{CasimirEPL}, colloidal mixtures under gravity\cite{kruppa:134106}, or even more far-reaching problems such as the growth of cancer cells\cite{chauviere:011210}.\\ 
The central dynamic variable within DDFT is the time-dependent density field $\rho(\mathbf{r},t)$. However, recent studies show that, based on Percus' test particle limit\cite{PhysRevLett.8.462}, it is also possible to use the DDFT equations 
to calculate dynamic correlation functions. Specifically, one may identify the van Hove dynamic correlation function with one-body density distributions of a mixture.\cite{hopkins2010,PhysRevE.75.040501} 
The equivalent function of the dynamic correlation function $G(\mathbf{r},t)$ in momentum space is the dynamic structure factor $S(\mathbf{k},t)$. Thus, besides calculating the dynamics of the density field one may use DDFT to gain more structural information about a phase-separating process. 
Here we investigate, in particular, the time-dependence of the average domain size. We also compare the DDFT predictions to those from a simpler, mean-field like (``Vineyard'') approach. \\
The rest of the paper is organized as follows. Section II contains the formulation of the model. After introducing the DDFT approach to the model in Sec.~III~A we provide more theoretical background for the linearized theory (Sec.~III~B) and the dynamical test particle theory (Sec.~III~C). In Sec.~IV~A, we study the conditions under which the system is unstable against spatio-temporal density perturbations by linearizing the DDFT equations.
For the nonlinear regimes, we provide numerical solutions of the (full) DDFT equations in Sec.~IV~B where we consider different dynamic variables, that is, the density fields $\rho_\alpha(\mathbf{r},t)$ (with $\alpha$ being the species index), the van Hove dynamic correlation functions $G_\alpha(\mathbf{r},t)$, 
and the resulting dynamic structure factor. 
We conclude in Sec.~V with a discussion of our results.

\section{Model}
Following our study in Ref.~\onlinecite{lichtner:024502}, we consider a system of overdamped Brownian particles consisting of two different species, where one species is magnetic $(m)$ and the other one is non-magnetic $(n)$. 
The cores are assumed to be ``soft'' for both species, such that two particles can penetrate each other given that the bulk density is sufficiently large. As a model for the core-core interactions we use the Gaussian Core Model\cite{stillinger:3968,PhysRevLett.85.2522,Likos2001267} (GCM), that is, $V^\text{core}(|\mathbf{r}-\mathbf{r'}|)=\varepsilon\exp\left[-(\mathbf{r}-\mathbf{r'})^2/\sigma^2\right]$, where $\mathbf{r}$ is the position coordinate on the $(x,z)$-plane
 and $\sigma$ roughly corresponds to the radius of gyration of the 'particles'. 
For the core potentials we set $\varepsilon^*=\varepsilon/(k_BT)>0$.
The (purely repulsive) interaction between pairs of non-magnetic particles is then given as $V_{nn}=V_{nm}=V^\text{core}(|\mathbf{r}-\mathbf{r'}|)$.
The particles from the magnetic species carry additional magnetic moments, such that the interaction between two magnetic particles is the sum of the soft core part and the spin part
\begin{align}
V_{mm}=V^\text{core}(|\mathbf{r}-\mathbf{r'}|)+V^\text{spin}(|\mathbf{r}-\mathbf{r'}|,\omega,\omega'),\label{eq.vcore}
\end{align}
where $\omega$ is a set of Euler angles representing the orientation of the three-dimensional unit spin vector $\mathbf{s}$.
For the spin-spin interactions we use a Heisenberg model
\begin{align}
V^\text{spin}=J(|\mathbf{r}-\mathbf{r'}|)\mathbf{s}\cdot\mathbf{s'},\label{eq.vspin}
\end{align}
where the range dependency is given by Yukawa's potential, that is, $J(|\mathbf{r}-\mathbf{r'}|)=-J\sigma\exp(-|\mathbf{r}-\mathbf{r'}|/\sigma-1)/|\mathbf{r}-\mathbf{r'}|$.
As in our previous study\cite{lichtner:024502}, we make the choice $J^*=J/(k_BT)>0$ such that ferromagnetic ordering is favored. We also note that we set $J(|\mathbf{r}-\mathbf{r'}|)=0$ for distances $|\mathbf{r}-\mathbf{r'}|<\sigma$, i.e. we assume that 
at these separations the interaction between two magnetic particles is negligible as compared to the repulsion from the core potentials. 

\section{dynamical density functional theory approach}
In classical density functional theory (DFT) for anisotropic particles the central quantity is the one-body profile $\rho(\mathbf{r},\omega)$. 
In our previous work\cite{lichtner:024502}, we introduced two \textit{time-dependent} one-body profiles $\rho_{m}(\mathbf{r},\omega,t)$, $\rho_{n}(\mathbf{r},\omega,t)$ representing both species at a given time $t$. 
The focus of Ref.~\onlinecite{lichtner:024502} was the investigation of nucleation phenomena. In the following, we will use a dynamical extension of classical DFT called ``Dynamical Density Functional Theory''\cite{marconi:a413,marconi:8032,ArcherSpinDec,AR04} (DDFT) in order to address the \textit{coarsening dynamics} of the phase-separating system.

We begin with the discussion of the theoretical framework for calculating $\rho_\alpha(\mathbf{r},\omega,t)$ in Sec.~\ref{secDDFTapp}, followed by a linear stability analysis of the key equations in Sec.~\ref{eq.lsa}. In Sec.~\ref{sec.dtpt} 
we extend our DDFT approach by observing the dynamics of a single (``self") particle in the ``sea'' of the remaining particles. 
This is a further step towards revealing
the underlying density-density correlations in the system and opens up access to van Hove's dynamic correlation functions $G_{\alpha\beta}(\mathbf{r},t)$.\cite{HansenSimpleLiquids3ed}

\subsection{Key ingredients}\label{secDDFTapp}
Our study is based on DDFT for the calculation of the \emph{time-dependent} one-body density profiles $\rho_{\alpha}(\mathbf{r},\omega,t)$ [with $\alpha=\{m,n\}$]. 
Within DDFT the dynamics is assumed to
be overdamped, i.e. inertial effects are neglected. The key approximation of DDFT is that the non-equilibrium two-body density distribution functions at time $t$ are set equal to those of an {\em equilibrium} system with the same one-body density profile.\\
The generalization of the DDFT approach for the binary mixture 
leads to two coupled integro-differential equations for the density profiles $\rho_\alpha(\mathbf{r},\omega,t)$,\cite{PhysRevE.80.021409,RexLoewen}
\begin{align}
\frac{\partial \rho_\alpha(\mathbf{r},\omega,t)}{\partial t}=&D\nabla \cdot\left[\rho_\alpha(\mathbf{r},\omega,t)\nabla \frac{\delta \mathcal{F}[\{\rho_\alpha(\mathbf{r},\omega,t)\}]}{\delta \rho_\alpha(\mathbf{r},\omega,t)}\right]\nonumber\\
&+D_r\hat{R}\cdot\left[\rho_\alpha(\mathbf{r},\omega,t)\hat{R}\frac{\delta \mathcal{F}[\{\rho_\alpha(\mathbf{r},\omega,t)\}]}{\delta \rho_\alpha(\mathbf{r},\omega,t)}\right].\label{eq:DDFTmixture}
\end{align}
In Eq.~(\ref{eq:DDFTmixture}), $\hat{R}=\omega\times\nabla_{\omega}$ is the rotation operator, the coefficients $D$ and $D_r$ are the translational and the rotational diffusion constants, respectively, and
$\delta \mathcal{F}[\rho]/\delta \rho$ is the functional derivative of the Helmholtz free energy functional, $\mathcal{F}$, with respect to the one-body density. 
The latter can be factorized into a translational number density part,
 $\rho_\alpha(\mathbf{r},t)$, and an orientational distribution function, $h_\alpha(\mathbf{r},\omega,t)$, that is,\cite{PhysRevE.58.3426}
\begin{align}
 \rho_\alpha(\mathbf{r},\omega,t)=\rho_\alpha(\mathbf{r},t)h_\alpha(\mathbf{r},\omega,t)\label{dftrho}
\end{align}	
with the normalization $\int d\omega\ h_\alpha(\mathbf{r},\omega,t)=1$. For the non-magnetic species, the orientational distribution function is homogeneous over the angular space, i.e. $h_n=1/4\pi$ for the non-magnetic species. 
In contrast, the orientational distribution of the magnetic species can become non-trivial as a result of a phase transition (or an external field).

The integro-differential equations for the overdamped dynamics of each species [see Eqs.~\eqref{eq:DDFTmixture}] are coupled via the Helmholtz free energy functional. The latter can be written as a sum of the ideal gas part and the excess (over ideal gas) part\cite{evansDFT}, that is, $\mathcal{F}=\mathcal{F}_\text{id}+\mathcal{F}_{\text{ex}}$.
Specifically, the contribution from the ideal gas is given by
\begin{align}
\mathcal{F}_\text{id}[\{\rho_\alpha\}]=k_BT&\sum\limits_{\alpha}^{\{m,n\}}\negthickspace\int\negthickspace d\mathbf{r}\negthickspace\int\negthickspace d\omega\rho_\alpha(\mathbf{r},\omega,t)\nonumber\\
&\times[\ln(\rho_\alpha(\mathbf{r},\omega,t)\Lambda_\alpha^2)-1],\label{eq.fidgas}
\end{align}
where $\Lambda_\alpha$ denotes the thermal de Broglie wavelength of species $\alpha$. For the excess free energy, we use a simple mean-field ansatz where the two-body density distribution is approximated by $\rho^{(2)}_{\alpha\beta}(\mathbf{r},\mathbf{r'},\omega,\omega',t)=\rho_\alpha(\mathbf{r},\omega,t)\rho_\beta(\mathbf{r'},\omega',t)$, that is, 
\begin{align}
\mathcal{F}_{\text{ex}}[\{\rho_\alpha\}]=\frac12\sum\limits_{\alpha,\beta}^{\{m,n\}}&\int\negthickspace d\mathbf{r}\negthickspace\int\negthickspace d\mathbf{r'}\negthickspace\int\negthickspace d\omega \negthickspace\int\negthickspace d\omega'\rho_\alpha(\mathbf{r,\omega},t)\nonumber\\
&\times V_{\alpha\beta}(|\mathbf{r}-\mathbf{r'}|,\omega,\omega')\rho_\beta(\mathbf{r'},\omega',t).\label{eq.fexc}
\end{align}

Following our previous work \cite{lichtner:024502}, we reduce the degrees of freedom in the present study by assuming that the magnetic moment relaxes instantaneously. This argument approximates situations where the orientational degrees of freedom relax much faster than the translational ones. As a consequence, we can \textit{approximately} set the functional derivative $\delta\mathcal{F}/\delta h_m(\mathbf{r},\omega,t)=0$ at all times $t$.
This yields a self-consistency relation for the orientational distribution $h_m(\mathbf{r},\omega,t)$ [see Ref.~\onlinecite{lichtner:024502} for details] 
\begin{align}
 h_m(\mathbf{r},\omega,t)=\dfrac{\exp(\mathbf{B}(\mathbf{r},t)\cdot \mathbf{s}(\omega,t))}{\int d\omega \exp(\mathbf{B}(\mathbf{r},t)\cdot \mathbf{s}(\omega,t))},\label{ddftangle}
\end{align}
where $\mathbf{s}$ is a three-dimensional normalized classical spin whose orientation is described by the Euler angles $\omega=(\theta,\varphi)$. In Eq.~(\ref{ddftangle}) the (self-consistent) effective field is given by
\begin{align}
\mathbf{B}(\mathbf{r},t)= -\negthickspace\int\negthickspace d\mathbf{r}'\negthickspace\int\negthickspace d\omega'\rho_m(\mathbf{r}',t)h_m(\mathbf{r}',\omega',t)J(|\mathbf{r}-\mathbf{r}'|)\mathbf{s'}\label{eq.efffield}.
\end{align}
Note that $\mathbf{B}(\mathbf{r},t)$ is always positive for the ferromagnetic coupling as we have included a negative sign in the definition of $J(|\mathbf{r}-\mathbf{r}'|)$ [see text below Eq.~(\ref{eq.vspin})].
This is an exact result within the mean-field approximation, which has been previously applied also to three-dimensional Heisenberg fluids\cite{PhysRevE.52.1915, PhysRevE.55.7242,PhysRevE.58.3426} as well as in other contexts such as in liquid crystal theory\cite{PhysRevA.38.2022}.
The local magnetization is given by the Langevin function $L(\mathbf{r},t)=\coth B(\mathbf{r},t)-1/B(\mathbf{r},t)$, which is identical to the angular averaged orientational distribution, that is,
\begin{align}
    m(\mathbf{r},t)=\int d\omega h_m(\mathbf{r},\omega,t)\cos\theta\equiv L(\mathbf{r},t).\label{eq.magnetiz}
\end{align}
In Eq.~(\ref{eq.magnetiz}), we assumed that the system has uniaxial symmetry around the direction $\mathbf{n}$ and $\cos\theta$ is the scalar product between the normalized vectors $\mathbf{s}$ and $\mathbf{n}$.
Our approximation of instantaneously relaxing orientational degrees of freedom implies that the last term on the right side of Eq.~(\ref{eq:DDFTmixture}) is negligible. The dynamical equations for the binary fluid thus reduce to
\begin{align}
 \frac{\partial \rho_\alpha(\mathbf{r},t)}{\partial t}=D\nabla\cdot \left[\rho_\alpha(\mathbf{r},t)\nabla \frac{\delta \mathcal{F}[\rho_n(\mathbf{r},t),\rho_m(\mathbf{r},t)]}{\delta \rho_\alpha(\mathbf{r},t)}\right],\label{eq.ddftbinaryrho}
\end{align}
combined with the self-consistency relation given by Eq.~(\ref{eq.efffield}). In the following, we measure time in units of the Brownian time scale $\tau_B =
\sigma^2/D$, which is of the order of $10^{-9}$s for typical Brownian particles.

\subsection{Linear stability analysis}\label{eq.lsa}


In the absence of an external potential ($V_{\rm ext}=0$) and under thermodynamically stable conditions, the stationary solution of Eqs.~(\ref{eq.ddftbinaryrho}) 
corresponds to homogeneous number densities, $\rho_\alpha^{0}$.
However, for phase separating systems
 homogeneous solutions are unstable inside the coexistence region.\cite{archer2011nucleation,BinderSD08,lichtner:024502} 
 In the following we aim to describe the \textit{onset} of this instability. To this end, we investigate the stability against spatio-temporal perturbations up to linear order. Following previous studies\cite{PhysRevE.85.061408} we consider small harmonic perturbations 
where it is assumed that the growth rate $\gamma(k)$ is the same for both species, that is,
\begin{align}
\rho_n(\mathbf{r},t)=&\rho_n^0+\Delta \rho(\mathbf{r},t)= \rho_n^0+\phi e^{i\mathbf{k}\cdot\mathbf{r}}e^{\gamma(k) t},\nonumber\\
\rho_m(\mathbf{r},t)=&\rho_m^0+\psi\Delta \rho(\mathbf{r},t)= \rho_m^0+ \psi\phi e^{i\mathbf{k}\cdot\mathbf{r}}e^{\gamma(k) t}.\label{ansatzpert}
\end{align}
Here, $\Delta \rho(\mathbf{r},t)$ is a small density perturbation where $\phi$ is the amplitude and $|\mathbf{k}|=k$ is the wave number.
We use $\psi$ as the ratio for the perturbation amplitudes between the non-magnetic species and the magnetic species. 
To describe the dynamics of $\Delta\rho(\mathbf{r},t)$, we approximate Eqs.~(\ref{eq.ddftbinaryrho}) by using a truncated Taylor series expansion of the functional derivatives of the free energy in $\Delta\rho(\mathbf{r},t)$, that is,
\begin{align}
 &\frac{\delta \mathcal{F}[\rho_n,\rho_m]}{\delta \rho_\alpha}=\left.\frac{\delta \mathcal{F}[\rho_n,\rho_m]}{\delta \rho_\alpha}\right\vert_{\rho_n^0,\rho_m^0}\nonumber\\
 &+\int d\mathbf{r'}\left.\frac{\delta^2 \mathcal{F}[\rho_n,\rho_m]}{\delta\rho_\alpha\delta\rho_\alpha}\right\vert_{\rho_n^0,\rho_m^0}(1-\delta_{\alpha m})\Delta\rho(\mathbf{r'},t)\nonumber\\
 &+\psi\int d\mathbf{r'}\left[\left.\frac{\delta^2 \mathcal{F}[\rho_n,\rho_m]}{\delta\rho_\alpha\delta\rho_\alpha}\right\vert_{\rho_n^0,\rho_m^0}\delta_{\alpha m}\right.\nonumber\\
 &\hspace{2cm}\left.+\left.\frac{\delta^2 \mathcal{F}[\rho_n,\rho_m]}{\delta\rho_\alpha\delta\rho_\beta}\right\vert_{\rho_n^0,\rho_m^0}(1-\delta_{\alpha m})\right]\Delta\rho(\mathbf{r'},t)\nonumber\\
 &+\int d\mathbf{r'}\left.\frac{\delta^2 \mathcal{F}[\rho_n,\rho_m]}{\delta\rho_\alpha\delta\rho_\beta}\right\vert_{\rho_n^0,\rho_m^0}\delta_{\alpha m}\Delta\rho(\mathbf{r'},t).\;\;\;\;(\beta\neq\alpha)\label{eq.taylor1}
\end{align}
In Eq.~(\ref{eq.taylor1}), $\delta_{\alpha m}$ is Kronecker's delta representing $1$ if $\alpha=m$ and $0$ otherwise.
Since we are interested in the \textit{local} stability of an equilibrium solution $\rho_\alpha^0$,
it is sufficient to truncate the expansion in Eq.~(\ref{eq.taylor1}) after the linear term. 
Inserting the resulting terms into Eqs.~(\ref{eq.ddftbinaryrho}) yields two coupled differential equations representing each species. For the non-magnetic species we obtain
\begin{align}
 &\frac{\partial \Delta\rho(\mathbf{r},t)}{\partial t}=\Gamma\vec\nabla\cdot\vec\nabla\Bigg[\Delta\rho(\mathbf{r},t)\nonumber\\
 &\quad+\rho_n^0\int d\mathbf{r'}\left.\dfrac{\delta^2 \mathcal{F}_{\text{ex}}[\rho_n,\rho_m]}{\delta\rho_n\delta\rho_n}\right\vert_{\rho_n^0,\rho_m^0}\Delta\rho(\mathbf{r'},t)\nonumber\\
 &\quad+\left.\psi\rho_n^0\int d\mathbf{r'}\left.\dfrac{\delta^2 \mathcal{F}_{\text{ex}}[\rho_n,\rho_m]}{\delta\rho_n\delta\rho_m}\right\vert_{\rho_n^0,\rho_m^0}\Delta\rho(\mathbf{r'},t)\right]\label{eq.lsanonmag}
\end{align}
and for the magnetic species,
\begin{align}
 &\psi\frac{\partial \Delta\rho(\mathbf{r},t)}{\partial t}=\Gamma\vec\nabla\cdot\vec\nabla\Bigg[\psi\Delta\rho(\mathbf{r},t)\nonumber\\
 &\quad+\psi\rho_m^0\int d\mathbf{r'}\left.\left.\dfrac{\delta^2 \mathcal{F}_{\text{ex}}[\rho_n,\rho_m]}{\delta\rho_m\delta\rho_m}\right\vert_{\rho_n^0,\rho_m^0}\Delta\rho(\mathbf{r'},t)\nonumber\right.\\
 &\quad+\;\;\;\left.\rho_m^0\int d\mathbf{r'}\left.\dfrac{\delta^2 \mathcal{F}_{\text{ex}}[\rho_n,\rho_m]}{\delta\rho_m\delta\rho_n}\right\vert_{\rho_n^0,\rho_m^0}\Delta\rho(\mathbf{r'},t)\right],\label{eq.lsamag}
\end{align}
where $\Gamma=D/(k_BT)$.
At this point we introduce the pair direct correlation functions $c^{(2)}_{\alpha\beta}(|\mathbf{r}-\mathbf{r'}|;\rho_n^0,\rho_m^0)$, defined as
\begin{align}
 k_BTc^{(2)}_{\alpha\beta}(|\mathbf{r}-\mathbf{r'}|;\rho_n^0,\rho_m^0)&=k_BTc^{(2)}_{\alpha\beta}(\mathbf{r},\mathbf{r'})\nonumber\\
 &=-\dfrac{\delta^2\mathcal{F}_\mathrm{ex}}{\delta\rho_\alpha(\mathbf{r'})\delta\rho_\beta(\mathbf{r})}.
\end{align}
It is then helpful to perform a Fourier transform with respect to the position coordinates. 
Since we are expanding around a homogeneous state, the correlation functions $c^{(2)}_{\alpha\beta}(k;\rho_n^0,\rho_m^0)$ depend only on the magnitude of $k$, and the same
is assumed to be true for the function $\gamma(k)$. This yields Eqs.~(\ref{eq.lsanonmag}) and (\ref{eq.lsamag}) in momentum space
\begin{align}
\gamma(k) \Delta\rho(\mathbf{k},t)=&-k^2\Gamma\Delta\rho(\mathbf{k},t)\left[1-\rho_n^0 c^{(2)}_{nn}(k;\rho_n^0,\rho_m^0)\right.\nonumber\\
&\hspace{2.4cm}\left.-\psi\rho_n^0 c^{(2)}_{nm}(k;\rho_n^0,\rho_m^0)\right]\nonumber\\
\psi\gamma(k)\Delta\rho(\mathbf{k},t)=&-k^2\Gamma\Delta\rho(\mathbf{k},t)\left[\psi-\psi\rho_m^0 c^{(2)}_{mm}(k;\rho_n^0,\rho_m^0)\right.\nonumber\\
&\hspace{2.4cm}\left.-\rho_m^0 c^{(2)}_{mn}(k;\rho_n^0,\rho_m^0)\right]\label{eq.linddft}.
\end{align}
We search for the solution $\gamma(k)$, which satisfies Eqs.~(\ref{eq.linddft}) simultaneously. To this end, we rewrite both equations in a matrix representation, that is,
\begin{align}
 \gamma(k)\begin{pmatrix} 1 \\ \psi \end{pmatrix} =\underline{\underline{M}}\cdot \underline{\underline{G}} \begin{pmatrix} 1 \\ \psi \end{pmatrix}.\label{eq.betamat}
\end{align}
Here, the matrices $\underline{\underline{M}}$ and $\underline{\underline{G}}$ are of dimension $2\times 2$. 
It follows that
\begin{align}
\underline{\underline{M}}&=\begin{pmatrix} -k^2\Gamma & 0 \\ 0 & -k^2\Gamma\end{pmatrix},\\
\underline{\underline{G}}&=\begin{pmatrix} 1-c^{(2)}_{nn}\rho_n^0 & -c^{(2)}_{nm}\rho_n^0\\ -c^{(2)}_{mn}\rho_m^0 & 1-c^{(2)}_{mm}\rho_m^0 \end{pmatrix}.\label{eq.gmat}
\end{align}
Since $\underline{\underline{M}}$ is diagonal and all diagonal elements are non-zero, the inverse $\underline{\underline{M}}^{-1}$ exists and the solution of Eqs.~(\ref{eq.betamat})-(\ref{eq.gmat}) reads
\begin{align}
 \gamma(k)=&\dfrac{\rm{Tr}(\underline{\underline{M}}\cdot\underline{\underline{G}})}{2}\pm \sqrt{\dfrac{\rm{Tr}(\underline{\underline{M}}\cdot \underline{\underline{G}})^2}{4}-\rm{det}(\underline{\underline{M}}\cdot \underline{\underline{G}})}\nonumber\\
 =&\frac{k^2\Gamma^2}{2}(c^{(2)}_{mm}\rho_m^0+c^{(2)}_{nn}\rho_n^0-2)\nonumber\\
 &\pm\frac{k^2\Gamma^2}{2}\left[(c^{(2)}_{nn})^2(\rho_n^0)^2+(c^{(2)}_{mm})^2\left(\rho_m^0\right)^2\right.\nonumber\\
 &\left.+4c^{(2)}_{nm}\rho_n^0c^{(2)}_{mn}\rho_m^0-2c^{(2)}_{mm}\rho_m^0c^{(2)}_{nn}\rho_n^0\right]^\frac12.\label{eq.gammaeq}
\end{align}
Our ansatz in Eq.~(\ref{ansatzpert}) shows that, for positive values of $\gamma(k)$, the density perturbation $\Delta\rho$ with wave number $k$ grows exponentially in time. 
Therefore, we search for the region
$k_<<k^*<k_>$ where $\gamma(k^*)>0$. Since $\gamma(k_<)=\gamma(k_>)=0$, the wave numbers $k_<$ and $k_>$ mark the transition points where the homogeneous fluid is linearly unstable ($\gamma(k)>0$) or linearly stable ($\gamma(k)<0$).

\subsection{Dynamical test particle theory}\label{sec.dtpt}
We now turn to the calculation of dynamical correlation functions. To this end we recall that, within the static DFT, there are two routes towards the calculation of the partial pair correlation functions for the homogeneous fluid\cite{PhysRevLett.85.1934}: 
the first one is the integral equation theory\cite{PhysRevE.70.031201,HansenSimpleLiquids3ed} based on the Ornstein-Zernike equation supplemented by an appropriate closure relation. The second one is the test particle method\cite{PhysRevLett.8.462}, where one particle of a given species is fixed at the origin and the partial pair correlation functions can be obtained from the one-body density profiles of the resulting inhomogeneous fluid.\cite{PhysRevLett.85.1934}
Here, we employ the dynamical extension\cite{hopkins2010,PhysRevE.75.040501} of the latter method, where the test particle is allowed to move away from the origin where it was at $t=0$.

To begin with, we define the relevant spatio-temporal correlation functions for the present system. 
The probability of finding a particle of species $\alpha$ at time $t$ at position $\mathbf{r}$, given that one particle of species $\beta$ was at the origin at time $t = 0$, is characterized by the van Hove functions $G_{\alpha\beta}(\mathbf{r}, t)$. 
Similar as for monodisperse suspensions, $G_{\alpha\beta}$ for a binary mixture can be decomposed into its respective ``self'' and ``distinct'' parts:\cite{hopkins2010,HansenSimpleLiquids3ed} 
\begin{align}
    G_{\alpha\beta}(\mathbf{r},t)=&G_{\alpha\beta}^s(\mathbf{r},t)+G_{\alpha\beta}^d(\mathbf{r},t),\label{eq.vanhovefct}
\end{align}
where
\begin{align}
    &G_{\alpha\beta}^s(\mathbf{r},t)=\frac{\delta_{\alpha\beta}}{N_\alpha}\left\langle \sum\limits_{i=1}^{N_\alpha} \delta\left(\mathbf{r}-\mathbf{r}^\alpha_i(t)+\mathbf{r}^\alpha_i(0)\right)\right\rangle,\nonumber\\
    &G_{\alpha\beta}^d(\mathbf{r},t)=\frac{1-\delta_{\alpha\beta}}{\sqrt{N_\alpha N_\beta}}\left\langle \sum\limits_{i=1}^{N_\alpha}\sum\limits_{j\neq i}^{N_\beta} \delta(\mathbf{r}-\mathbf{r}^\alpha_i(t)+\mathbf{r}^\beta_j(0))\right\rangle\nonumber\\
    &\hspace{1.6cm}+\delta_{\alpha\beta} \frac{1}{N_\alpha}\left\langle \sum\limits_{i=1}^{N_\alpha}\sum\limits_{j\neq i} \delta\left(\mathbf{r}-\mathbf{r}^\alpha_i(t)+\mathbf{r}^\alpha_j(0)\right)\right\rangle,
\end{align}
with $\delta_{\alpha\beta}$ being Kronecker's delta. Further, $N=N_\alpha+N_\beta$ is the total number of particles in the system. At time $t=0$ these functions fulfill the initial conditions 
\begin{align}
G_{\alpha\beta}^s(\mathbf{r},t=0)&=\delta_{\alpha\beta}\delta(\mathbf{r}),\nonumber\\
G_{\alpha\beta}^d(\mathbf{r},t=0)&=\rho_\alpha^\text{bulk} g_{\alpha\beta}(\mathbf{r}),\label{eq.boundary}
\end{align}
where $g_{\alpha\beta}(\mathbf{r})$ denotes the partial (static) pair distribution functions. 

As explained above, one can find the equilibrium structure of a fluid from Percus' test particle limit.\cite{PhysRevLett.8.462} 
In the following, we apply an ``identification scheme'' together with Percus' test particle route suggested by Refs.~\onlinecite{hopkins2010,PhysRevE.75.040501} in order to obtain the time-dependent (off-equilibrium) van Hove functions $G_{\alpha\beta}(\mathbf{r},t)$. 
To this end, we identify the self part and the distinct part of van Hove's function with conditional one-body profiles of the fluid mixture, where one particle is treated seperately from the rest. 
The dynamics of the conditional profiles is then given by the DDFT equations of a \textit{four-component} fluid mixture, where both of the original components ($\alpha=\{m,n\}$) are decomposed into a 
self part ($s$) and a distinct part ($d$),
\begin{align}
 \frac{\partial \rho_\alpha^i(\mathbf{r},t)}{\partial t}=D\nabla\cdot \left[\rho_\alpha^i(\mathbf{r},t)\nabla \frac{\delta \mathcal{F}[\{\rho_\alpha^i(\mathbf{r},t)\}]}{\delta \rho_\alpha^i(\mathbf{r},t)}\right],\quad i=\{s,d\}.\label{eq.ddftvhove}
\end{align}
The self part refers to a single particle and the distinct part to the remaining 
$N_\alpha-1$ particles. 
Further, the free energy functional entering Eqs.~(\ref{eq.ddftvhove}) is given as $\mathcal{F}=\mathcal{F}_\text{id}+\mathcal{F}_{\text{ex}}$ where 
we have (as a generalization of the conventional mixture)
\begin{align}
\mathcal{F}_\text{id}[\{\rho_\alpha^i\}]=&k_BT\sum\limits_{i}^{\{s,d\}}\sum\limits_{\alpha}^{\{m,n\}}\negthickspace\int\negthickspace d\mathbf{r}\negthickspace\int\negthickspace d\omega\rho^i_\alpha(\mathbf{r},\omega,t)\nonumber\\
&\quad\times[\ln(\rho^i_\alpha(\mathbf{r},\omega,t)\Lambda_\alpha^2)-1],\nonumber\\
\mathcal{F}_{\text{ex}}[\{\rho^i_\alpha\}]=&\frac14\sum\limits_{i,j}^{\{s,d\}}\sum\limits_{\alpha,\beta}^{\{m,n\}}\int\negthickspace d\mathbf{r}\negthickspace\int\negthickspace d\mathbf{r'}\negthickspace\int\negthickspace d\omega \negthickspace\int\negthickspace d\omega'\rho^i_\alpha(\mathbf{r,\omega},t)\nonumber\\
&\quad\times V^{ij}_{\alpha\beta}(|\mathbf{r}-\mathbf{r'}|,\omega,\omega')\rho^j_\beta(\mathbf{r'},\omega',t).\label{eq.vhoveenergy}
\end{align}
The interactions $V_{\alpha\beta}^{ij}$ entering Eqs.~(\ref{eq.vhoveenergy}) are determined by the pair potentials $V_{\alpha\beta}$ as defined in Eqs.~(\ref{eq.vcore}-\ref{eq.vspin}). Specifically, we set $V_{\alpha\beta}^{dd}=V_{\alpha\beta}^{sd}=V_{\alpha\beta}$ 
where 
\begin{align}
V_{nn}&=V_{mn}=V_{nm}=V^\text{core}(|\mathbf{r}-\mathbf{r'}|),\nonumber\\
V_{mm}&=V^\text{core}(|\mathbf{r}-\mathbf{r'}|)+J(|\mathbf{r}-\mathbf{r'}|)\mathbf{s}\cdot\mathbf{s'}.
\end{align}
Moreover, to take into account the fact that there is only one test particle in the system, we set $V_{\alpha\beta}^{ss}=0$ for all $\alpha,\beta$  
where $V_{\alpha\alpha}^{ss}=0$ takes the absence of self-interactions into account.
The self-consistent solution for the effective field is given by
\begin{align}
\mathbf{B}(\mathbf{r},t)= -\negthickspace\int\negthickspace d\mathbf{r}'\negthickspace&\int\negthickspace d\omega'[\rho_m^s(\mathbf{r}',t)+\rho_m^d(\mathbf{r}',t)]\nonumber\\
&\times h_m(\mathbf{r}',\omega',t)J(|\mathbf{r}-\mathbf{r}'|)\mathbf{s'}\label{eq.efffieldvhove}.
\end{align}
In order to suffice the initial conditions given by Eqs.~(\ref{eq.boundary}) we evolve the one-body profiles $\rho_{\alpha}^i(\mathbf{r},t)$ via Eqs.~(\ref{eq.ddftvhove}) [together with Eqs.~(\ref{eq.vhoveenergy}-\ref{eq.efffieldvhove})] while holding the position of the test particle fixed. 
This ``relaxation'' procedure yields the static pair correlation functions. 
After this initial preparation we release the test particle and evolve the one-body profiles forward in time which gives the \textit{time-dependent} van Hove correlation functions via the identification scheme
\begin{align}
G_{\alpha\beta}^s(\mathbf{r},t)=\rho_{\beta}^s(\mathbf{r},t),\;\; \mathrm{and}\;\; G_{\alpha\beta}^d(\mathbf{r},t)=\rho_{\alpha}^d(\mathbf{r},t).\label{eq.tpident} 
\end{align}
Note that we consider a single test particle of species $\alpha$ surrounded by the remaining fluid, i.e. we set $\rho_\beta^s(\mathbf{r},\omega,t)=0$ (with $\beta\neq\alpha$) for all times $t$.
This reduces Eqs.~(\ref{eq.ddftvhove}) to three coupled equations.
Furthermore, we note that in contrast to Eq.~(\ref{eq.efffield}) the effective field in Eq.~(\ref{eq.efffieldvhove}) is given by the ``full'' (magnetic) van Hove function $G_{m\alpha}(\mathbf{r},t)$ where $\alpha$ is a test particle either from the magnetic \textit{or} from the non-magnetic species.
This result can be obtained by using for $G^s_{m\alpha}(\mathbf{r},t)$ and $G^d_{m\alpha}(\mathbf{r},t)$ the same ansatz Eq.~(\ref{ddftangle}) for the orientational distribution function, i.e. we set $h_m^s(\mathbf{r},\omega,t)=h_m^d(\mathbf{r},\omega,t)$. 
Minimizing then Eq.~(\ref{eq.vhoveenergy}) with respect to the effective field $B(\mathbf{r},t$) yields Eq.~(\ref{eq.efffieldvhove}).






\begin{figure}[tpb]
\includegraphics[width=9cm]{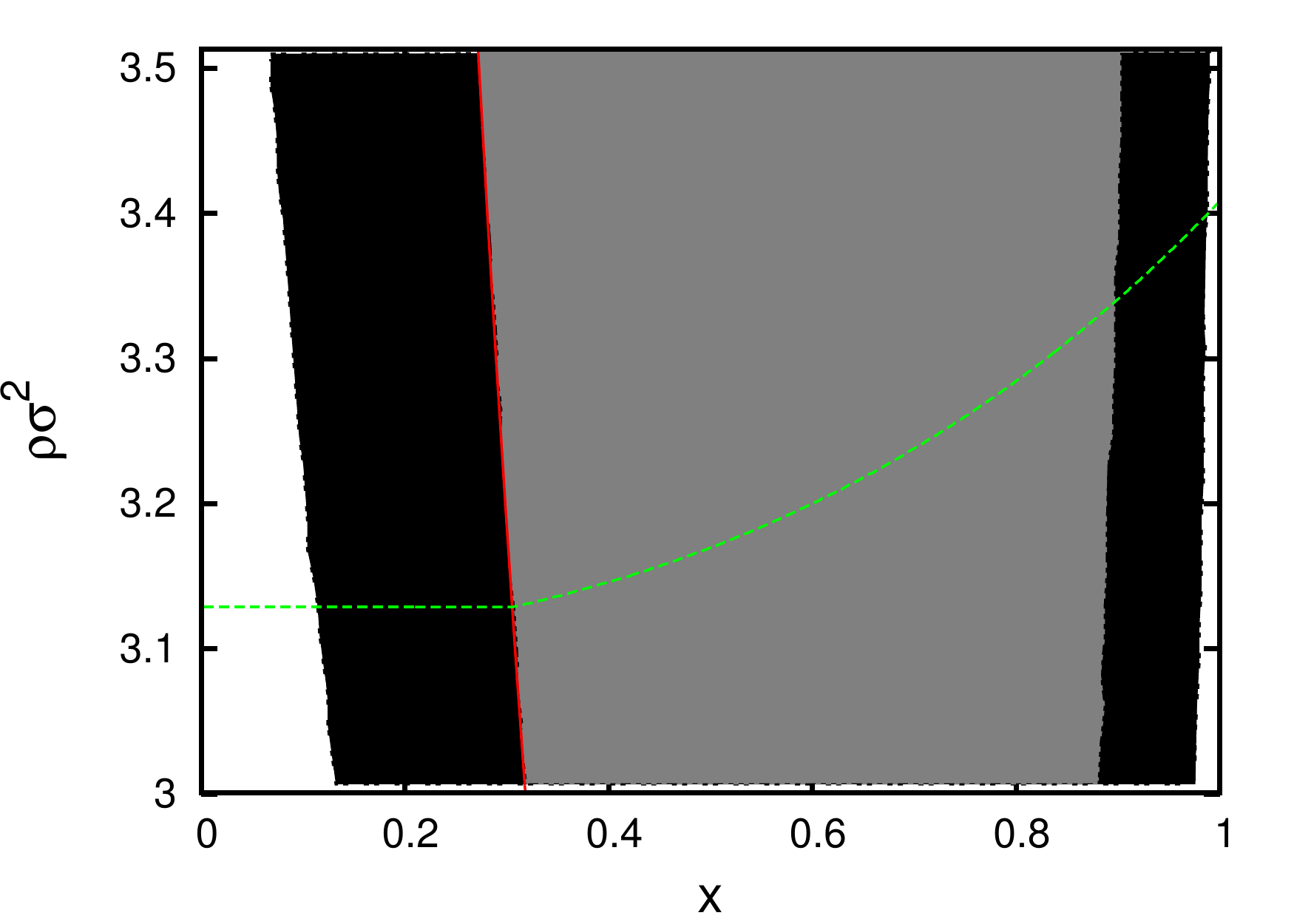}
\caption{(Color online) The phase diagram for the bulk binary (non-magnetic/magnetic) mixture in the density-concentration plane, where $x$ is the concentration of the magnetic species. The coupling parameters are $\varepsilon^*=5.0$, $J^*=0.5$. The areas shown in black and grey are the metastable region and the unstable region, respectively. The green dashed line is the $P^*=P\sigma^2/(k_BT)=80$ isobar. The Curie line is indicated by the red curve.}
\label{fig.phasediag}
\end{figure}

\section{Results}
\subsection{Equilibrium phase behavior and linear stability}
We concisely recall the phase behaviour of the bulk binary mixture in the density-concentration plane (see Fig.~\ref{fig.phasediag}; for details see Ref.~\onlinecite{lichtner:024502}). 
Depending on the bulk density of the system we find for coupling parameters $J/\varepsilon>0$ [where $J$ is the ferromagnetic, $\varepsilon$ the repulsive coupling constant, respectively] a demixing phase transition 
driven by the ferromagnetic interactions. 
The latter point is seen from the fact that the demixing is always coupled to a transition from a paramagnetic phase rich in $n$-particles to a ferromagnetic phase rich in $m$-particles.
The Curie line (shown as a red curve in Fig.~\ref{fig.phasediag}) separates all unmagnetized from all magnetized equilibrium states.
To identify coexisting states from Fig.~\ref{fig.phasediag}, we recall such states are characterized by equal pressure. As an example, we have included the isobar with pressure $P^*=P\sigma^2/(k_BT)=80$ as a green dashed curve in Fig.~\ref{fig.phasediag}. 
Depending on the distance of the concentration $x$ to the coexistence curve (binodal), the state considered is either metastable (shown as the black areas in Fig.~\ref{fig.phasediag}) or unstable (shown as the grey area that is bounded by the spinodal). 
Both regions are obtained by solving the thermodynamic stability equations for the binary mixture in an isothermal-isobaric ensemble (see Ref.~\onlinecite{lichtner:024502}).\\
In the following we choose a fixed repulsion strength $\varepsilon^*=5.0$ and a fixed magnetic coupling strength $J^*=0.5$. The first parameter is well below the ``freezing'' limit of the GCM repulsion strength $\varepsilon_u$. 
Indeed, the corresponding freezing temperature of the GCM in three dimensions is $t_u=k_BT/\varepsilon_u\approx 0.008$, i.e. $\varepsilon_u/(k_BT)\approx 125$.
At $\varepsilon^*=5$, the system therefore remains fluid at all densities.\cite{LLWLoewen2000} Furthermore, the coupling parameter $J^*$ is chosen such that we find demixing states for a broad range of concentrations $x$ as well as for high density values.\\
\begin{figure*}[tpb]
 \includegraphics[height=5.8cm]{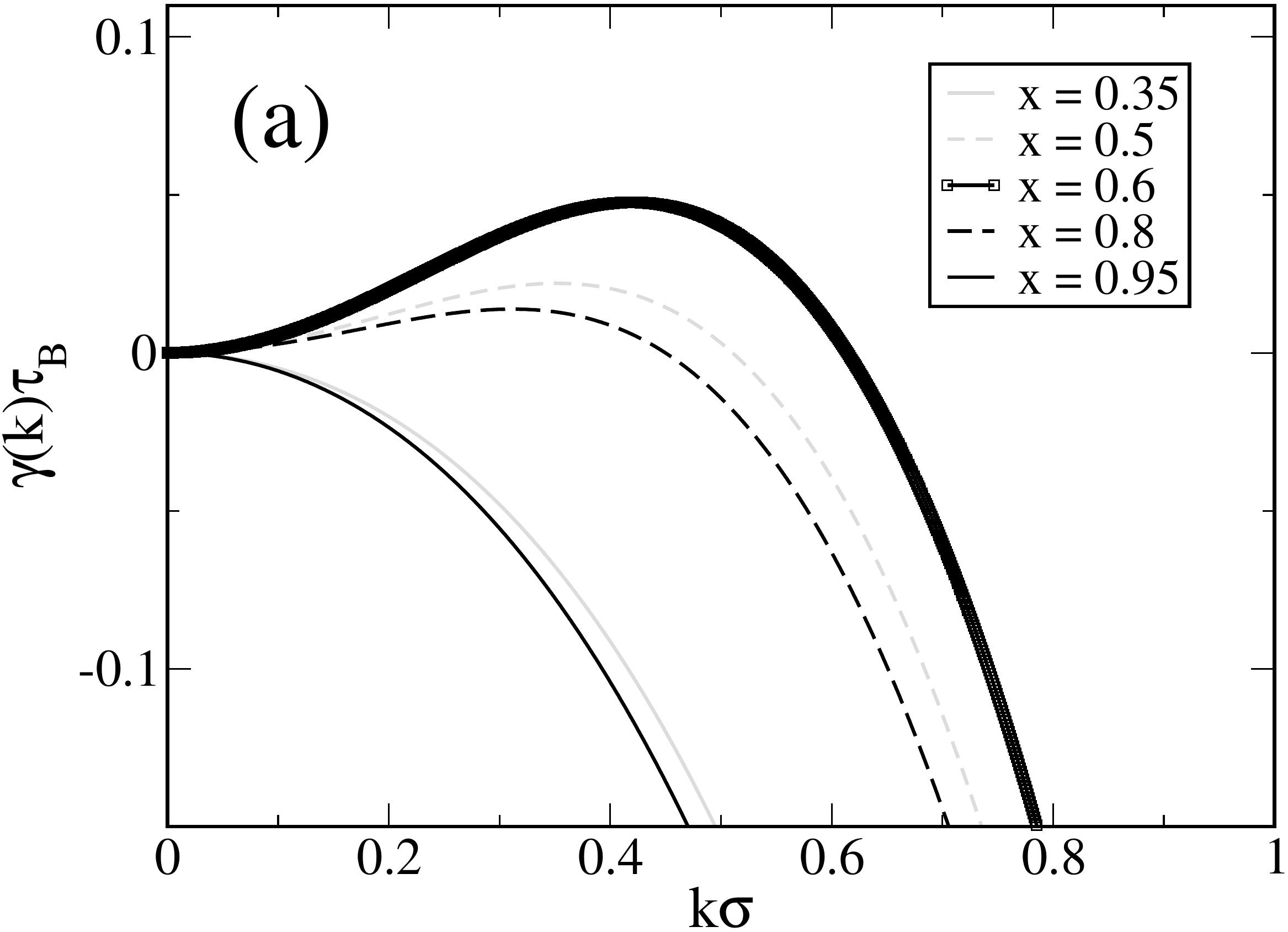} \hspace{0.5cm}\includegraphics[height=6cm]{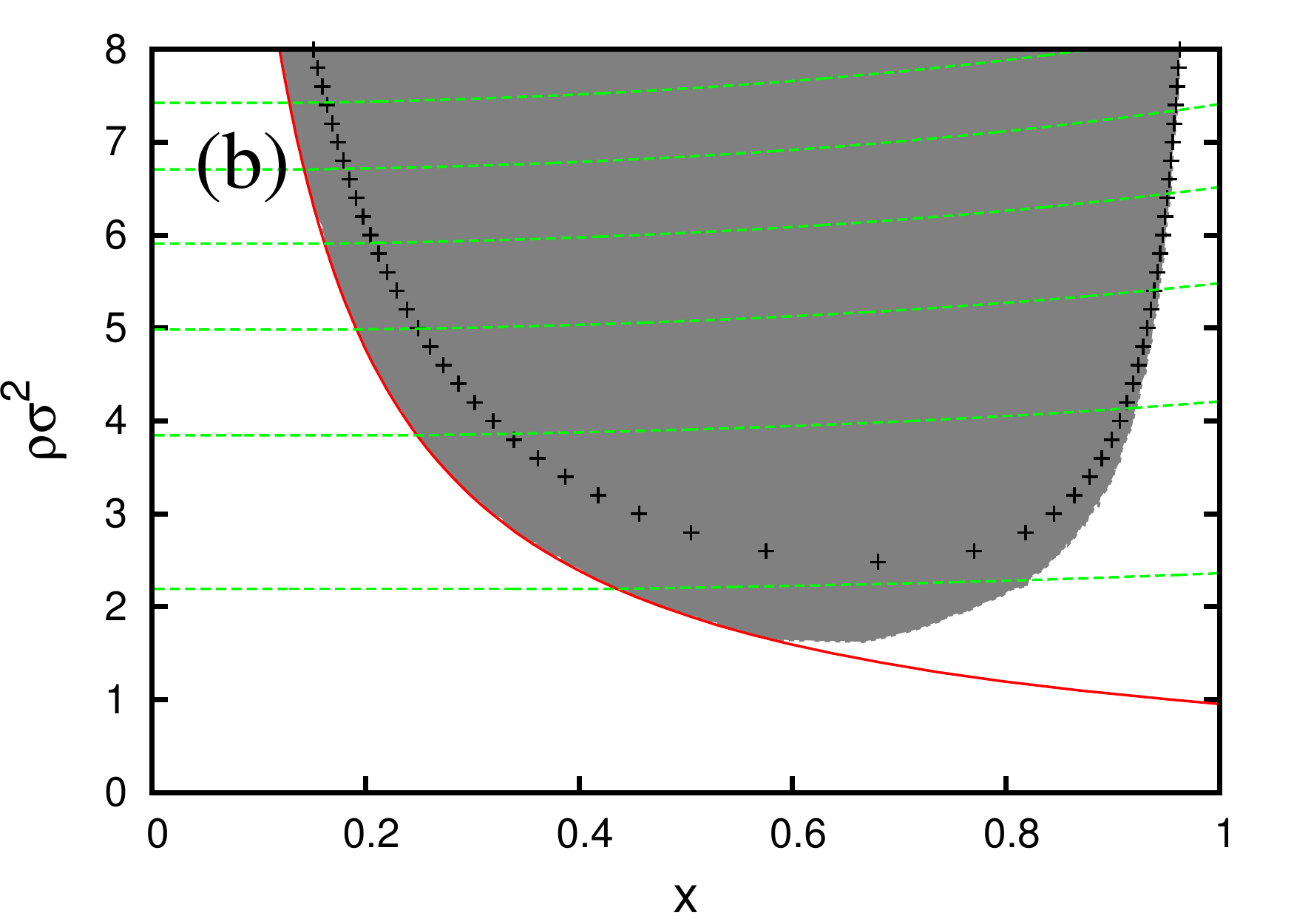}
\caption{(Color online) (a) shows results for the growth rate $\gamma(k)$ obtained from Eq.~(\ref{eq.gammaeq}) at density $\rho\sigma^2=3.2$ at different concentrations within the meta- and unstable part of the phase diagram.
(b) shows the unstable region of the phase diagram. The black crosses mark the crossover region where solutions $\gamma(k)>0$ can be found within the linear analysis. 
The green lines are isobars with a line-to-line pressure difference of $\Delta P^*=80$ (where the lowest curve is for $P^*=40$). The Curie line is indicated by the red curve. 
The parameters are $\varepsilon^*=5$ and $J^*=0.5$.}
\label{fig.lengthscale1}
\end{figure*}
With this background, we now discuss the linear stability against spatio-temporal perturbations based on numerical solution of  
Eq.~(\ref{eq.gammaeq}). Specifically, we choose a number density of $\rho\sigma^2=3.2$ as it is known that the mean-field approximation is particularly accurate at high densities, i.e., in situations where the average number of next neighbors is large for each particle.\cite{PhysRevE.62.7961,PhysRevE.64.041501} 
We focus on concentrations $x$ inside the unstable region of the phase diagram (grey area in Fig.~\ref{fig.phasediag}). For all concentrations $x$ considered the growth rate is $\gamma\approx 0$ for $k\rightarrow 0$ (corresponding to large wave lengths $\lambda=2\pi/k$) as can be seen in Fig.~\ref{fig.lengthscale1}(a).
For concentrations $0.42< x< 0.86$, we find solutions $\gamma>0$ for $k\neq 0$ and $k<k_>$ where
the critical wavenumber $k_>$ depends on the concentration $x$. 
All $\gamma$-curves  exhibit local maxima (e.g., $k_{\text{max}}\sigma\approx 0.42$ for $x=0.6$), i.e. density perturbations with wave number $k_{\text{max}}$ are expected to grow the fastest.
By increasing the wave number further such that $k\geq k_>(x)$, the growth rate becomes negative for all $k$.\\
We note that the predictions from linear stability analysis become inconsistent with the phase diagram near the spinodal [boundary of the grey area in Figs.~\ref{fig.phasediag} and \ref{fig.lengthscale1}(b)]. 
This becomes clear, when we calculate the region where we find critical wavenumbers $\left.k_>(x)\right\vert_\rho\neq 0$ for different number densities $\rho$. The black crosses in Fig.~\ref{fig.lengthscale1}(b) mark the crossover region 
where $\left.k_>(x)\right\vert_\rho \rightarrow 0$, i.e. outside this region the linear stability analysis predicts that spinodal decomposition does not occur. However, from Fig.~\ref{fig.lengthscale1}(b) we see that this region is always \textit{inside} the spinodal; thus, the size of the unstable region is underestimated. 
Only for higher densities the black crosses seem to approach the spinodal (we checked values up to $\rho\sigma^2=12$).\\
The fact that the linear analysis becomes wrong close to the spinodal is also confirmed by our numerical solution of 
the full DDFT equations (including the nonlinear terms): These numerical calculations reveal spontaneous demixing to occur even close to the spinodal.\\
A similar conclusion regarding the performance of a linear analysis close to the spinodal was reported in a study\cite{dhont:5112} based on the nonlinear Smoluchowksi equation (which is closely related to DDFT, see Ref.~\onlinecite{ArcherSpinDec}). 
There, it was argued that terms linear in $\Delta\rho$, which are related to the absolute stability conditions of the system and are therefore proportional to inverse susceptibilites, become 
small upon approach of the spinodal. Thus, the terms linear in $\Delta \rho$ are no longer dominant near the spinodal and nonlinear terms have to be taken into account. 
A further reason for the observed inconsistency between our linear stability analysis and the full DDFT calculations may be the approximate character of the direct correlation functions 
entering our theory. In fact, similar inconsistencies occur in the context of equilibrium properties\cite{HansenSimpleLiquids3ed}, although we should note that, at least for the pure GCM fluid, these inconsistencies are typically small\cite{PhysRevE.62.7961}.




\subsection{Spinodal decomposition kinetics}\label{SDkinetics}
\begin{figure}[tpb]
\includegraphics[width=8cm]{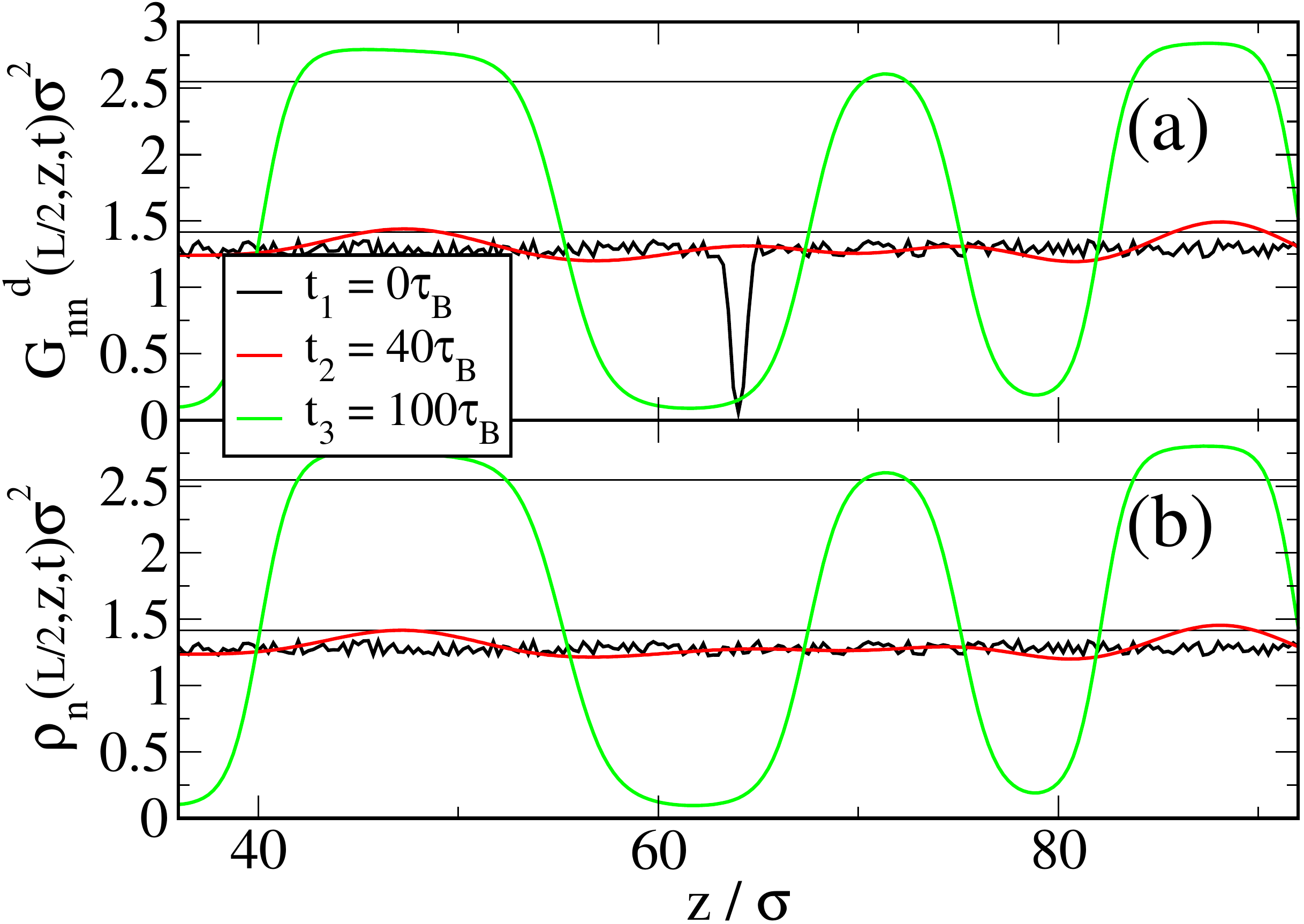}
\caption{(Color online) Time series for the phase separating system for a slice through $x=L/2$ (where $L=128\sigma$). The times are $t_1=0\tau_B$, $t_2=40\tau_B$ and $t_3=100\tau_B$. (a) shows the distinct part of the van Hove function for the non-magnetic species, where a non-magnetic test particle has been placed at position $x=z=64\sigma$ at time $t_1=0\tau_B$ leading to a correlation hole at the same position (see black curve). (b) shows the one-body density as a function of $z$-position for the same species. 
The black horizontal lines [plotted at the same values in (a) and (b)] are a guide to the eye. The coupling parameters are $\varepsilon^*=5.0$, $J^*=0.5$.}
\label{fig.rhovhove}
\end{figure}

\begin{figure*}[ht]
\hspace*{-1.5cm}
\begin{minipage}[b]{0.23\linewidth}
\centering
\resizebox{55mm}{!}{\includegraphics{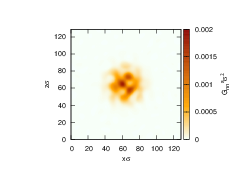}}
\end{minipage}
\hspace{0.2cm}
\begin{minipage}[b]{0.23\linewidth}
\centering
\resizebox{55mm}{!}{\includegraphics{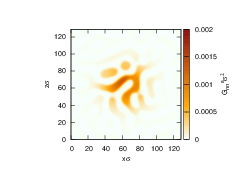}}
\end{minipage}
\hspace{0.2cm}
\begin{minipage}[b]{0.23\linewidth}
\centering
\resizebox{55mm}{!}{\includegraphics{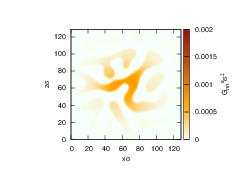}}
\end{minipage}
\hspace{0.1cm}
\begin{minipage}[b]{0.23\linewidth}
\centering
\resizebox{55mm}{!}{\includegraphics{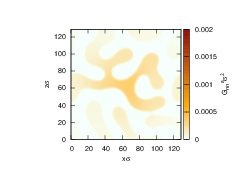}}
\end{minipage}\\
\hspace*{-1.5cm}
\begin{minipage}[b]{0.23\linewidth}
\centering
\resizebox{55mm}{!}{\includegraphics{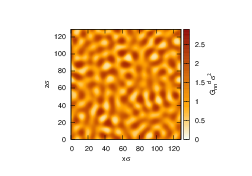}}
\end{minipage}
\hspace{0.2cm}
\begin{minipage}[b]{0.23\linewidth}
\centering
\resizebox{55mm}{!}{\includegraphics{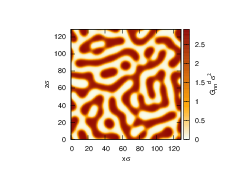}}
\end{minipage}
\hspace{0.2cm}
\begin{minipage}[b]{0.23\linewidth}
\centering
\resizebox{55mm}{!}{\includegraphics{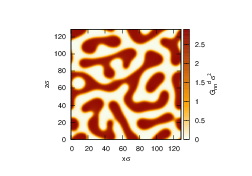}}
\end{minipage}
\hspace{0.1cm}
\begin{minipage}[b]{0.23\linewidth}
\centering
\resizebox{55mm}{!}{\includegraphics{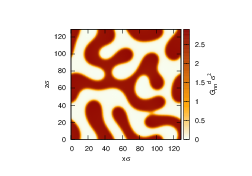}}
\end{minipage}\\
\hspace*{-1.5cm}
\begin{minipage}[b]{0.23\linewidth}
\centering
\resizebox{55mm}{!}{\includegraphics{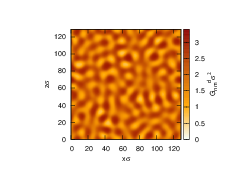}}
\end{minipage}
\hspace{0.2cm}
\begin{minipage}[b]{0.23\linewidth}
\centering
\resizebox{55mm}{!}{\includegraphics{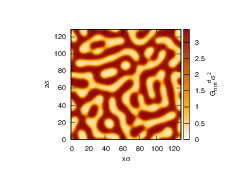}}
\end{minipage}
\hspace{0.2cm}
\begin{minipage}[b]{0.23\linewidth}
\centering
\resizebox{55mm}{!}{\includegraphics{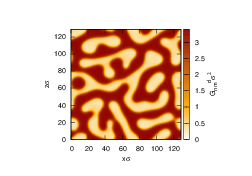}}
\end{minipage}
\hspace{0.1cm}
\begin{minipage}[b]{0.23\linewidth}
\centering
\resizebox{55mm}{!}{\includegraphics{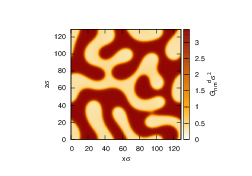}}
\end{minipage}
\caption{(Color online) Snapshots of the partial correlation functions $G_{\alpha\beta}^i(\mathbf{r},t)$ for the non-magnetic self part (upper row), the non-magnetic distinct part (middle row) and the magnetic distinct part (bottom row) as a function of the position. 
The non-magnetic test particle was inserted at the position $x=z=64\sigma$ at time $t=0$. The time increases from the left to the right: $t_1=60\tau_B$, $t_2=140\tau_B$, $t_3=280\tau_B$ and $t_4=800\tau_B$. The parameters are $\rho\sigma^2=3.2$, $x=0.6$, $\varepsilon^*=5.0$ and $J^*=0.5$.}
\label{fig.gfct}
\end{figure*}
For a fluid undergoing spinodal decomposition three different time regimes have to be distinguished.\cite{KapralBook} At early times the density fluctuations are small. Thus, on a theoretical description
level one may consider the terms linear in the density fluctuation, such as in the Cahn-Hilliard theory\cite{cahn:258,CahnHilliard,Cahn1961795}. For intermediate times the interfacial width $\xi$ plays an important role, but sharp interfaces have not been fully formed yet.
This does not happen until the late stages of the domain coarsening, where the ratio $\xi/l(t)$ [where $l(t)$ is the average domain size] becomes negligible small.\cite{KapralBook} 
A key feature of late stage domain growth is that the average domain size follows a power law behavior $l(t)\propto t^\delta$, where the exponent $\delta$ 
strongly depends on the system as well as on the nature of the order parameter\cite{KapralBook}: 
for non-conserved scalar order parameters where the late stage kinetics is driven by the interfacial curvature the exponent is given by $\delta=\frac12$; 
on the other hand, for conserved scalar order parameters the late stage domain growth is found to be slower with an exponent $\delta=\frac13$.
In the present system the demixing transition is coupled to the spatio-temporal changes of the one-body profiles $\rho_\alpha(\mathbf{r},t)$, which are \textit{conserved} quantities, i.e., the particle numbers
\begin{align}
 N_n=\int d\mathbf{r}\ \rho_n(\mathbf{r},t), \;\;\;  N_m=\int d\mathbf{r}\ \rho_m(\mathbf{r},t),
\end{align}
are constant for all times $t$. This is a general feature of the DDFT method. 
We stress that the magnetization profile $m(\mathbf{r},t)$ is coupled to the corresponding one-body profile $\rho_m(\mathbf{r},t)$ at any time instant [see Eq.~(\ref{eq.efffield})]. Thus, the cluster growth behaviour should be determined mainly by the conserved order parameters $\rho_m(\mathbf{r},t)$ and $\rho_n(\mathbf{r},t)$. 
In the following, we aim for a more detailled study of the demixing kinetics within our system focussing on the time-dependent average cluster size $l_\alpha(t)$. 
\subsubsection{Correlation functions}

Our main target quantities are the van Hove correlation functions $G_{\alpha\beta}(\mathbf{r},t)$, from which the average domain sizes $l_\alpha(t)$ can be calculated quite straight forwardly [see Eq.~(\ref{eq.kaverage}) below]. 
To illustrate the time-dependence of these functions we place exemplarily a non-magnetic test particle at position $x=z=L/2$ (centre position of the system) at time $t=0$. 
We note that we approximate the initial conditions given in Eqs.~(\ref{eq.boundary}) by preparing a ``correlation hole'' for the non-magnetic distinct part $G_{nn}^d$ such that $G_{nn}^d+G_{nn}^s=\text{const.}$ The constant is adjusted to a bulk density value where we expect spinodal decomposition to occur (e.g., $\rho\sigma^2=3.2$ at $x=0.6$, see Fig.~\ref{fig.phasediag}). After adding noise to each (partial) van Hove function $G_{\alpha\beta}^i$ we use Eqs.~(\ref{eq.ddftvhove}) to iterate the functions forward in time. \\
In Fig.~\ref{fig.rhovhove}(a) we show a time series for the distinct part $G_{nn}^d$ for a constant $x=L/2$-slice (where $L$ is the system size). 
For comparison we have also included results for the one-body density profile $\rho_n(L/2,z,t)$ obtained from solving Eqs.~(\ref{eq.ddftbinaryrho}) together with Eqs.~(\ref{eq.fexc}-\ref{eq.efffield}). 
At time $t=0$ the correlation hole in the region of $z=64\sigma(=L/2)$ corresponding to a very localized test particle can be clearly seen for the function $G_{nn}^d$ [see black curve in Fig.~\ref{fig.rhovhove}(a)]. For times $t\lesssim 20\tau_B$, this correlation
hole is being ``filled up'' again by neighboring particles since the released test particle can move on the substrate. 
By comparing the results of $G_{nn}^d$ with $\rho_n$ in Fig.~\ref{fig.rhovhove}, we can define a correlation time $t_c$ after which the local density-density correlations induced by the test particle (completely) disappear. 
Close inspection shows that for times $t< t_c\simeq 100\tau_B$ the interfaces are more pronounced for the function $G_{nn}^d$ [compare, e.g., the local maxima at $z=88\sigma$ for the red curves in Fig.~\ref{fig.rhovhove}]. 
For later times $t\geq t_c$, these differences become unnoticeable to the eye.\\ 
In order to further illustrate the density-density correlations in the surrounding region of the initial position of the test particle, we present snapshots for various partial van Hove functions $G_{\alpha\beta}^i$ in Fig.~\ref{fig.gfct}. It is seen that, after 
a time period of order $t_c\simeq 100\tau_B$, the spatial structures displayed by the self part $G_{nn}^s(\mathbf{r},t)$ [shown in the upper row of Fig.~\ref{fig.gfct}] are similarly extended in space as those of the distinct part $G_{nn}^d(\mathbf{r},t)$ (as well as with similar topology 
but with different absolute values since the test particle is normalized to $\int d\mathbf{r}\ G_{nn}^s(\mathbf{r},t)=1$).
This indicates that the spatio-temporal correlations between different particles became weak after the correlation time $t_c$. 
\begin{figure}[tpb]
\includegraphics[width=8cm]{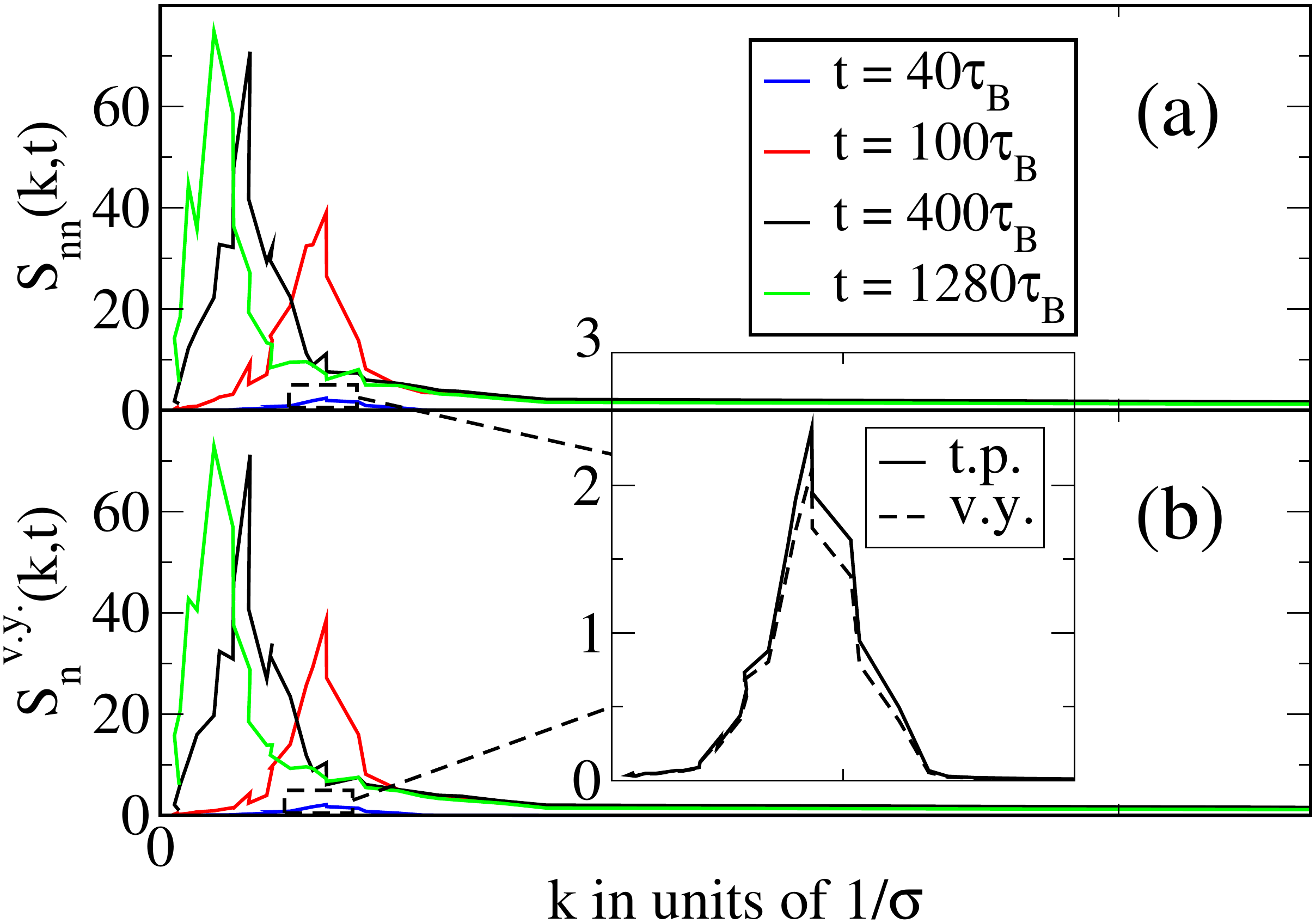}
\caption{(Color online) Dynamic structure factor $S(k,t)$ for the non-magnetic species as a function of the angular-averaged wave number $k$. (a) shows results obtained from the test particle method and the data in (b) is calculated via a Vineyard type approximation. The inset shows an enlarged view of the curves for the time $t=40\tau_B$.}
\label{fig.strfact_nonmag}
\end{figure}
At later times $t\gtrsim t_c$, one observes labyrinth structures with sharp boundaries between two species. This is the typical coarsening behavior characterizing late-stage spinodal decomposition.
We note that these structures are similar in character to those displayed by the individual density fields reported in our previous work.\cite{lichtner:024502}
\subsubsection{Domain size}
In the following, we aim to determine the average domain size $l_\alpha(t)$ for each species $\alpha$. There are several established routes for calculating the function $l_\alpha(t)$: first, one could determine $l_\alpha(t)$ from the first zero-crossing of $G_{\alpha\beta}(\mathbf{r},t)$. 
Second, one may obtain $l_\alpha(t)$ from the first moment of the dynamic structure factor. This is the route we are following here.\\
The dynamic structure factor corresponds to the Fourier transform of the (full) van Hove function [see Eq.~(\ref{eq.vanhovefct})], that is, 
\begin{align}
S_{\alpha\alpha}(\mathbf{k},t)=\int d\mathbf{r}e^{i\mathbf{k}\mathbf{r}}[G_{\alpha\alpha}^s(\mathbf{r},t)+G_{\alpha\alpha}^d(\mathbf{r},t)]. \label{eq.sfacttp}
\end{align}
Here we average the angle-dependent function $ S_{\alpha\alpha}(\mathbf{k},t)$ over all directions of $\mathbf{k}$ within the $x$-$y$-plane yielding the angle-averaged structure factor $ S_{\alpha\alpha}(k,t)$. 
The first moment of this function is given by
\begin{align}
\langle k_\alpha\rangle_t=\dfrac{\int k S_{\alpha\alpha}(k,t)dk}{\int S_{\alpha\alpha}(k,t)dk}.\label{eq.kaverage}
\end{align}
From that we can calculate the average domain size as $l_\alpha(t)=2\pi/\langle k_\alpha\rangle_t$.\\  
\begin{figure}[tpb]
\includegraphics[width=7.8cm]{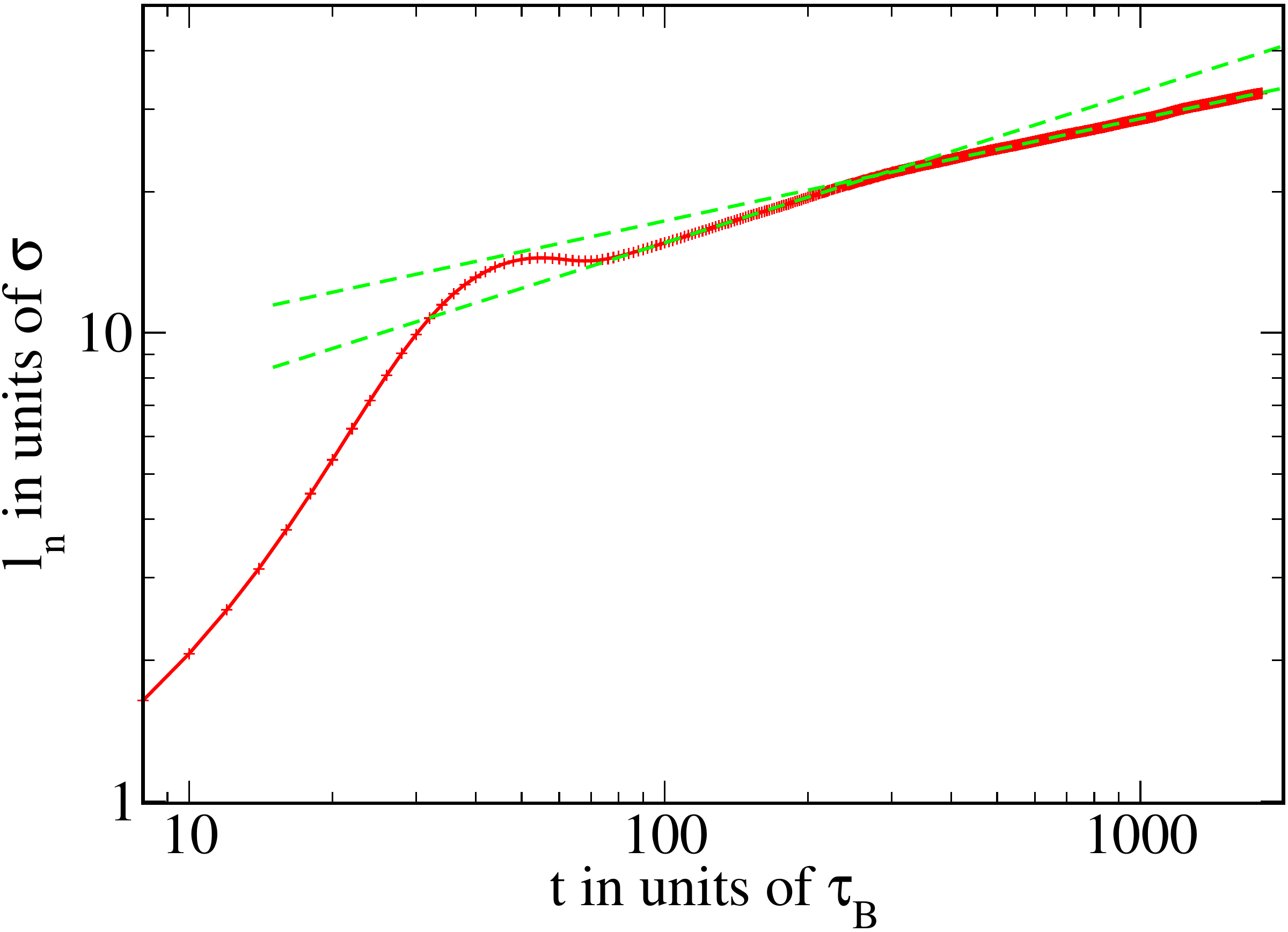}
\caption{(Color online) The average domain size $l_n(t)$ for the non-magnetic species as a function of time in a double logarithmic representation as obtained from Eq.~(\ref{eq.sfacttp}) via the test particle route. The fitting functions are shown as dashed lines. The coupling parameters are $\varepsilon^*=5.0$ and $J^*=0.5$.}
\label{fig.lengthscale_nonmag}
\end{figure}
In Fig.~\ref{fig.strfact_nonmag}(a) we show results for the (angle averaged) dynamic structure factor $S_{nn}(k,t)$ obtained from Eq.~(\ref{eq.sfacttp}) [choosing data for the non-magnetic species as an example]. 
Clearly, we find one dominating average cluster size for all considered times as can be seen from the single peak structure of $S_{nn}(k,t)$. As time increases, the peak is shifted towards smaller wave numbers corresponding to larger average cluster sizes in the system.\\ 
We now turn to the time dependence of the resulting domain sizes. 
For times $t\lesssim 100\tau_B$ we recall that sharp interfaces have still not fully evolved in both parts of the van Hove functions $G_{nn}^s(\mathbf{r},t)$, $G_{nn}^d(\mathbf{r},t)$ [cf. Figs.~\ref{fig.rhovhove}(a),\ref{fig.gfct}]. 
Consequently, we do not observe a power law behavior for $l_n(t)$ on this time scale as can be seen in Fig.~\ref{fig.lengthscale_nonmag}.
For later times $t\gtrsim 100\tau_B$, on the other hand, we find that the function follows a power law behavior $l_n(t)\propto t^{\delta_n}$ with exponent $\delta_n\simeq 0.323$ similar to the power-$1/3$ rule that is expected for systems with conserved order parameters.\cite{KapralBook} However, this exponent changes abruptly to $\delta_n\simeq 0.218$ for times $t\gtrsim 280\tau_B$.
We suspect that this crossover could be a consequence of a finite size effect: at $t\gtrsim 280\tau_B$ the domain growth has progressed so far that some domains penetrate the boundaries. 
We recall that we use periodic boundary conditions, i.e. any particle transport current going through the boundary reenters from the other side. Thus, each affected domain artificially separates into two ``distinct'' domains resulting in decreased values for $l_n(t)$.
\begin{figure}[tpb]
\includegraphics[width=8cm]{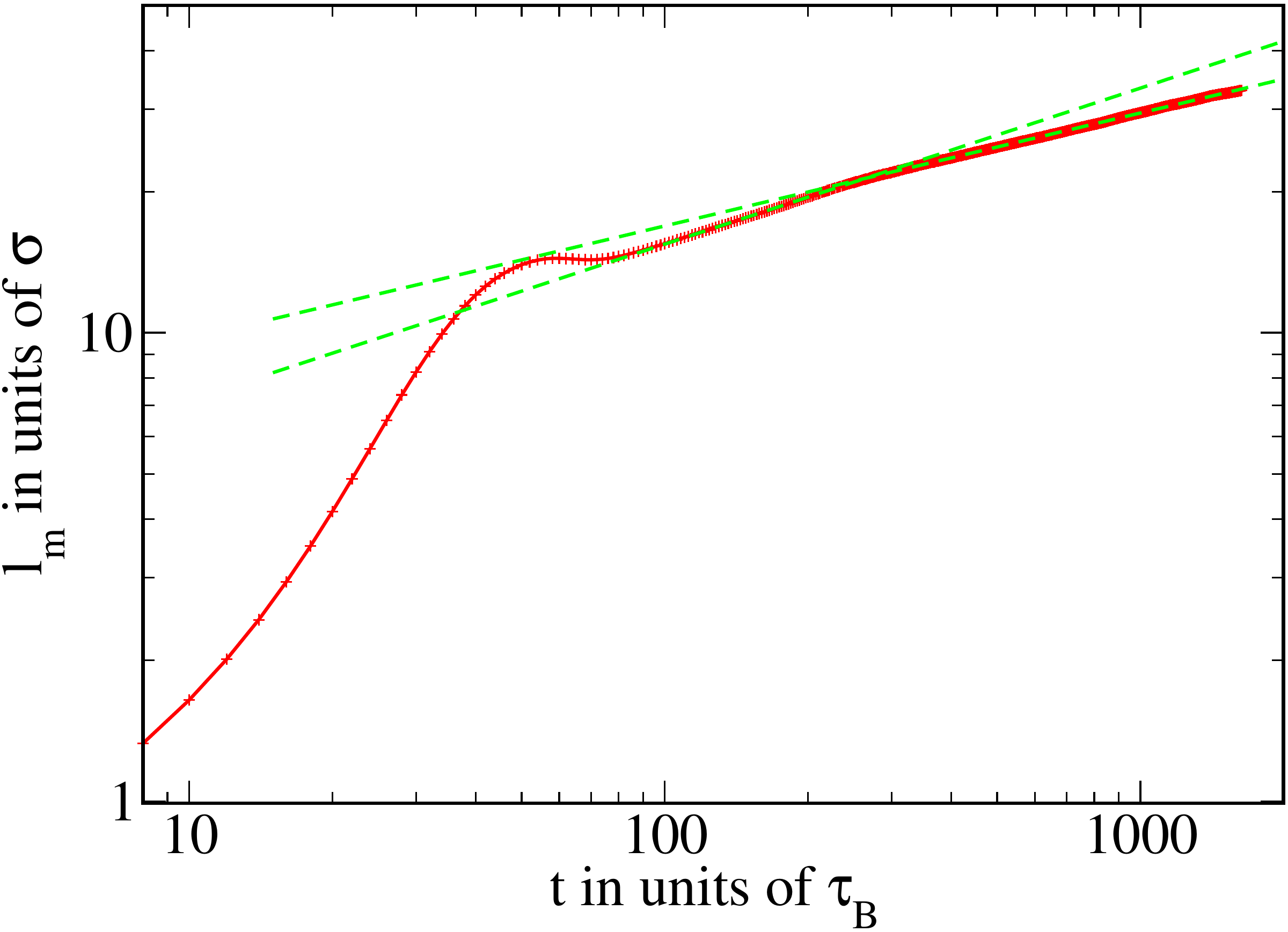}
\caption{(Color online) Same as Fig.~\ref{fig.lengthscale_nonmag}, but for the magnetic species. The fitting functions are shown as dashed lines. The coupling parameters are $\varepsilon^*=5.0$ and $J^*=0.5$.}
\label{fig.lengthscale_mag}
\end{figure}
Supporting our arguments, Brownian dynamics simulations for Lennard-Jones-type particles have shown that spinodal decomposition kinetics only follows the ``classical'' power law behavior with exponent $\delta=1/3$ (for conserved systems) 
as long as the average domain size does \textit{not} approach the scale of the simulation cell.\cite{lodge97,lodge98}\\ 
In Fig.~\ref{fig.lengthscale_mag} we show the corresponding results for the magnetic species. 
After sharp interfaces have been formed (cf. Figs.~\ref{fig.rhovhove},\ref{fig.gfct}), we observe a $t^{\delta_m}$ power law behavior for the characteristic length $l_m(t)$, 
where $\delta_m\simeq 0.333$. For times $t\gtrsim 280\tau_B$ this exponent changes to $\delta_m\simeq 0.241$. Hence, the general behavior for the growth of the average magnetic domain is consistent with our observations made for the non-magnetic species.\\
\begin{figure}[tpb]
\includegraphics[width=8cm]{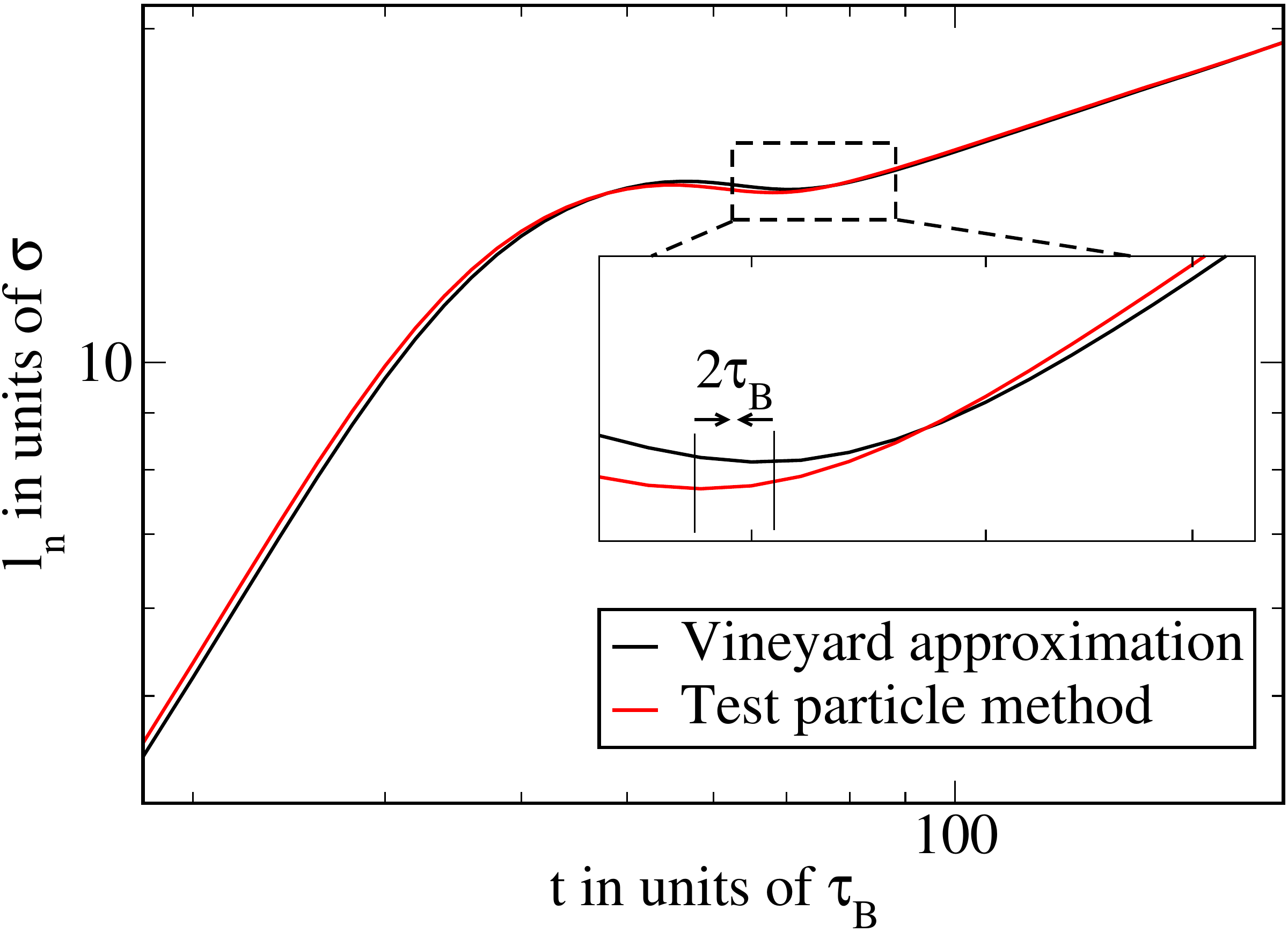}
\caption{(Color online) Log-log representation of the average domain size (non-magnetic species) as a function of time. The black curve is a result obtained from a Vineyard type approximation [calculated via the relation $l_n^{v.y.}=2\pi/\langle k_n\rangle_t^{v.y}$ using Eqs.~(\ref{eq.kaverage},\ref{eq.sfactor})] and the red curve is a test particle method result. 
The inset shows a close-up view of the range of times where the formation of sharp interfaces occurs [cf. Fig.~\ref{fig.gfct}].
The parameters are $\rho=3.2$, $\varepsilon^*=5$ and $J^*=0.5$.}
\label{fig.ltcomparison}
\end{figure}
Finally, it is interesting to compare the results based on the true dynamic structure factor (obtained from the van Hove function) with those from the so-called Vineyard approximation\cite{ArcherSpinDec,dhont:5112}. Within the latter, 
the dynamic structure factor is given by the absolute value of the one-body density profile $\rho_\alpha(\mathbf{r},t)$, that is,
\begin{align}
 &S_{\alpha}^{\text{v.y.}}(\mathbf{k},t)=\left\vert \hat{\rho}_\alpha(\mathbf{k},t) \right\vert,\nonumber\\
 &\hat{\rho}_\alpha(\mathbf{k},t)=\int d\mathbf{r}e^{i\mathbf{k}\mathbf{r}}\rho_\alpha(\mathbf{r},t),\ \ \alpha=\{m,n\}.\label{eq.sfactor}
\end{align}
Thus, the Vineyard approximation is a mean-field approximation which neglects spatio-temporal correlations. 
Typical results for the angle-averaged function $S_{\alpha}^{\text{v.y.}}(k,t)$ are shown in Fig.~\ref{fig.strfact_nonmag}(b). Comparing with those for the full structure factor, we find that 
for times $t>t_c\simeq 100\tau_B$ the results obtained from both approaches are very similar. On the other hand, for smaller times $t\lesssim t_c$ (where, according to the van Hove functions 
[see Figs.~\ref{fig.rhovhove}(a),\ref{fig.gfct}], sharp interfaces are yet not present) the differences become apparent. 
In order to further illustrate this, we show a comparison for the resulting average domain size $l_n(t)$ in Fig.~\ref{fig.ltcomparison}.  
Again, we find quantitative differences for the cluster growth for the time regime $t\lesssim t_c$. 
It seems that the spatio-temporal density-density correlations induced by inserting a test particle in the center of the system act as a catalyzer for the demixing transition: we find that the average domain growth in the test particle method is ``ahead'' in time by approximately $2\tau_B$ compared to the simple Vineyard type approximation.
After $t_c\simeq 100\tau_B$ the Vineyard type approximation produces similar results compared to the results from the test particle approach. 
This is again a sign that all (initial) density-density correlations have disappeared
after the correlation time $t_c$.

\section{conclusions}
In this study we have employed DDFT to investigate the coarsening dynamics of a binary colloidal system where one species carries an additional magnetic moment. We 
considered the bulk fluid mixture quenched inside the coexistence region. 
After presenting the bulk phase diagram (see also Ref.~\onlinecite{lichtner:024502}), we first performed a linear stability analysis to identify 
those wave numbers $k$ that correspond to \textit{growing} harmonic density perturbations. 
We find that the linear analysis underestimates the occurence of spinodal decomposition resulting in wrong predictions for states near the spinodal. Indeed, other studies\cite{dhont:5112} confirm that in this case nonlinear terms have to be included for the equation of motion.  
Moreover, we note that DDFT is able to show spontaneous demixing for the full unstable region of the phase diagram being thermodynamically consistent in this regard.\\
In the second part, we used a DDFT approach to address the \textit{real-time} coarsening dynamics during spinodal decomposition. 
By inserting a test particle into the homogeneous fluid combined 
with an ``identification scheme'' suggested by Refs.~\onlinecite{hopkins2010,PhysRevE.75.040501} [which in essence is an extension of Percus' test particle method towards relaxation dynamics] we obtained the (partial) van Hove functions $G_{\alpha\beta}^i(\mathbf{r},t)$.
This route opened up an access to calculate the average cluster size $l_\alpha(t)$. We showed that the coarsening during the first-order demixing transition is characterized by different time-scales: 
for times $t\lesssim20\tau_B$ we find that the diffusion-controlled relaxational dynamics of the test particle in the free energy landscape formed by the ``sea'' of the remaining particles is predominant. 
After this time scale the formation of sharp domain interfaces takes place while the average cluster size $l_\alpha$ increases. 
Within a time period of approximately $100\tau_B$ we observe the formation of sharp interfaces throughout the system. 
For times beyond this correlation time we find 
late-stage spinodal decomposition where the average cluster growth (as obtained from the dynamic structure factor) is given by a power-law behaviour $l_\alpha\propto t^{\delta_\alpha}$ with $\delta_n\simeq 0.323$ and $\delta_m\simeq 0.333$ being the exponents for the non-magnetic species
and the magnetic species, respectively. We note that these exponents are in agreement with the ``classical'' power law for conserved systems, that is, $\delta_\text{cl}=1/3$.\cite{KapralBook}\\
We also compared the predictions for $S_{\alpha\alpha}(k,t)$ and $l_\alpha(t)$ from the test particle scheme with corresponding ones from the simpler Vineyard approximation. The latter neglects spatio-temporal 
density correlations. 
Moreover, the effective mean field $B(\mathbf{r},t)$ for the angular distribution function $h_m(\mathbf{r}',\omega',t)$ is linear in $\rho_m(\mathbf{r},t)$ within the Vineyard approximation. 
In contrast, in Eq.~(\ref{eq.efffieldvhove}) we use the full magnetic van Hove correlation function $G_{mm}(\mathbf{r},t)=G_{mm}^s(\mathbf{r},t)+G_{mm}^d(\mathbf{r},t)$ for the calculation of $B(\mathbf{r},t)$. 
By comparing the test particle results to the Vineyard (mean-field) results we conclude that the early time regime of spinodal decomposition where density-density correlations play a role is finished after a correlation time of $t_c\simeq 100\tau_B$.
Furthermore, we find that local density-density correlations in the vicinity of the test particle position tend to support the coarsening process during spinodal decomposition: for times up to $t_c$ the interfacial structures are more pronounced as compared to the Vineyard system. 
For very late times 
these differences fade, i.e. the average cluster growth $l_\alpha(t)$ for both approaches is similar for $t\gtrsim t_c$.\\
In the present paper we considered only one (high) bulk density value of $\rho\sigma^2=3.2$, as it is known that the mean-field approximation for the (excess) Helmholtz free energy is quite accurate for ``soft'' systems where the number of next-neighbors is large. 
It would be very interesting to improve the present density functional approach beyond the mean-field level, e.g. using a modified mean-field approach (MMF)\cite{PhysRevE.58.3426}, or alternatively a perturbation expansion of the (excess) Helmholtz free energy as a further strategy. 
This would also open up pathways towards a deeper understanding of the impact of correlations at lower values of the bulk density.\\
Another interesting aspect concerns the interplay of the phase behavior and demixing dynamics of the 2D ``bulk'' system considered here with further external influences. 
In fact, a number of recent experiments\cite{PhysRevLett.100.148304,PhysRevLett.99.038303,PhysRevLett.109.070601,PhysRevLett.104.255703} 
have addressed this topic by exposing magnetic systems to patterned surfaces, and (additional) external magnetic fields. 
The present DDFT approach could be easily generalized to include such effects by adjusting the free energy functional. 
Work in these directions is under way.

\acknowledgements
We gratefully acknowledge financial support via the Collaborative Research Center (SFB) 910, ``Control of self-organizing nonlinear systems: Theoretical methods and concepts of application" and the Collaborative Research Center (SFB) 951 ``Hybrid Inorganic/Organic Systems for Opto-Electronics".

%

\end{document}